\documentclass[aps,prl,reprint,amsfonts,amssymb,amsmath,nofootinbib]{revtex4-2}

\usepackage[english]{babel}
\usepackage[utf8]{inputenc}
\usepackage[colorinlistoftodos, color=green!40, prependcaption]{todonotes}
\usepackage{amsthm}
\usepackage{mathtools}
\usepackage{physics}
\usepackage{xcolor}
\usepackage{graphicx}
\usepackage[left=23mm,right=13mm,top=35mm,columnsep=15pt]{geometry} 
\usepackage{adjustbox}
\usepackage{placeins}
\usepackage[T1]{fontenc}
\usepackage{lipsum}
\usepackage{csquotes}
\usepackage{subfiles}
\usepackage{hyperref}
\hypersetup{pdftitle={Article}, pdfauthor={Author}}
\usepackage{graphicx}                   
\usepackage[caption=false]{subfig}
\usepackage[left]{lineno}
\usepackage{cleveref} 
\usepackage[normalem]{ulem} 
\usepackage{siunitx}
\usepackage{booktabs}
\usepackage{caption}
\usepackage{titlesec}
\titleformat{\subsection}{\bfseries}{\thesubsection.}{1em}{}
\titlespacing*{\subsection}{0pt}{1.25ex plus 1ex minus .2ex}{0ex}

\Crefname{equation}{Eq.}{Eqs.}
\Crefname{figure}{Fig.}{Figs.}

\begin{document}

\title{Searching for scalar field dark matter with LIGO}

\author{Alexandre S. G\"ottel$^{1,\dag}$}
\author{Aldo Ejlli$^2$}
\author{Kanioar Karan$^2$}
\author{Sander M. Vermeulen$^3$}
\author{Lorenzo Aiello$^{4,5}$}
\author{Vivien Raymond$^1$}
\author{Hartmut Grote$^1$}

\affiliation{$^1$Gravity Exploration Institute, Cardiff University, Cardiff CF24 3AA, United Kingdom}
\affiliation{$^2$Max-Planck-Institute for Gravitational Physics and Leibniz University Hannover, Callinstr. 38, 30167 Hannover, Germany}
\affiliation{$^3$California Institute of Technology, Department of Physics, Pasadena, California 91125, USA}
\affiliation{$^4$Università di Roma Tor Vergata, I-00133 Roma, Italy}
\affiliation{$^5$INFN, Sezione di Roma Tor Vergata, I-00133 Roma, Italy}
\affiliation{$^\dag$gottela@cardiff.ac.uk}
\date{\today}

\begin{abstract}
We report on a direct search for scalar field dark matter using data from LIGO's third observing run.
We analyse the coupling of size oscillations of the interferometer's beamsplitter and arm test masses that may be caused by scalar field dark matter. Using new efficient search methods to maximise sensitivity for signatures of such oscillations, we set new upper limits for the coupling constants of scalar field dark matter as a function of its mass, which improve upon bounds from previous direct searches by several orders of magnitude in a frequency band from \SIrange{10}{180}{\hertz}.

\end{abstract}
\maketitle
\noindent Laser interferometers, with their extreme sensitivity to minute length changes, have revolutionized astronomy with a wide range of gravitational-wave (GW) detections over the last years~\cite{abbott2023}. Due to their capabilities at or beyond quantum limits, GW detectors can also be used directly in the search for new physics, without the mediation of gravitational waves, for example in the search for dark matter (DM). Several ideas have been put forward as to how different candidates of DM can directly couple to GW detectors, ranging from scalar field DM~\cite{stadnik2015,grote2019,Morisaki2019} to dark photon DM~\cite{pierce2018,Morisaki2021}, and to clumpy DM coupling gravitationally or through an additional Yukawa force~\cite{hall2018}. Scalar field DM influences objects by \emph{accelerating} them when a field gradient exists, and by \emph{expanding} them (and altering their refractive index) without net acceleration. Upper limits for scalar field DM have been set using data from the GEO\,600 GW detector~\cite{vermeulen2021a}, the Fermilab Holometer~\cite{Aiello2022}, and LIGO~\cite{Fukusumi2023}. The scalar field searches of the GEO\,600 and Holometer instruments used the dominant \emph{expansion} (and refractive index) effect in those instruments, while the work of \cite{Fukusumi2023} used the \emph{acceleration} effect. Likewise, upper limits on Dark photon DM, which is in simplified terms represented by a vector field causing objects to accelerate, have been set using data from the first (O1) and third (O3) observing runs of the LIGO  detectors~\cite{guo2019,DPDM2022}.

\noindent In this work, we analyse the \emph{expansion} effect of scalar field DM on dual-recycled Fabry-Pérot Michelson interferometers such as LIGO~\cite{Aasi_2015}, Virgo~\cite{Acernese_2015}, or KAGRA~\cite{10.1093/ptep/ptaa125}.
We also develop enhanced spectral search techniques that are optimised to search 
efficiently at a lower frequency (\textit{i.e.} mass) range of DM, and apply these to
search for DM signatures using data of the third observing run of the two LIGO observatories. Not finding viable candidates, we set new upper limits that surpass existing bounds by up to three orders of magnitude in a frequency band from \SIrange{10}{180}{\hertz}, and are competitive up to \SI{5000}{\hertz}.

\subsection*{Expected dark matter signal}
The astronomically-inferred mass density associated with DM may be attributed to an undiscovered scalar field with a high occupation number. Models of weakly-coupled low-mass ($\ll 1\, \mathrm{eV}$) scalar fields predict that sufficient particles could be produced in the early universe through a vacuum misalignment mechanism and manifest as the observed DM. This scalar field DM would behave as a coherently oscillating classical field~\cite{arvanitaki2015,stadnik2015}:
\begin{equation}\label{eq:osc_field}
        \phi(t,\vec{r}) = \phi_0 \cos\left(\omega_\phi t - \vec{k}_\phi \cdot \vec{r}\right),
\end{equation}
where $\omega_\phi = m_\phi c^2 / \hbar$ is the angular Compton frequency, $\vec{k}_\phi = m_\phi\vec{v}_{\text{obs}} / \hbar $ is the wave vector, $m_\phi$ is the mass of the field, and $\vec{v}_{\text{obs}}$ the velocity relative to the observer. The amplitude of the field can be set as $\phi_0 = \hbar \sqrt{2 \rho_{\mathrm{local}}} / (m_\phi c)$, under the assumption that this scalar field constitutes the local DM density $\rho_{\mathrm{local}}$ \cite{read2014}.  

Moreover, dynamical models of scalar field DM predict that such matter would be trapped and virialised in gravitational potentials, leading to a Maxwell-Boltzmann-like distribution of velocities $\vec{v}_{\text{obs}}$ relative to an observer. As non-zero velocities produce a Doppler-shift of the observed DM field frequency, this virialisation results in the DM field having a finite coherence time or, equivalently, a spread in observed frequency (linewidth) \cite{derevianko2018,pierce2018}. The expected linewidth is ${\Delta\omega_{\text{obs}}/\omega_{\text{obs}}\sim 10^{-6}}$ for DM trapped in the galactic gravity potential, as in the standard galactic DM halo model. Similarly, the observed frequency $\omega_{\text{obs}}$ shifts from $\omega_\phi$ by a $\propto v_\text{obs}^2$ term, negligible in our analysis.

Scalar field DM could couple to the fields of the Standard Model (SM) in various ways. These couplings are modelled by the addition of a parameterised interaction term to the SM Lagrangian~\cite{ringwald2012,hees2018}. In this paper, we consider linear interaction terms with the 
electromagnetic field tensor $F_{\mu\nu}$ and the electron rest mass $m_e$:
\begin{equation}\label{eq:L_int}
    \mathcal{L}_\mathrm{int} \supset \frac{\phi}{\Lambda_\gamma} \frac{F_{\mu\nu}F^{\mu\nu}}{4}  - \frac{\phi}{\Lambda_e} m_e \bar{\psi}_e \psi_e,
\end{equation}
where $\psi_e$, $\bar{\psi}_e$ are the SM electron field and its Dirac conjugate, respectively, and $\Lambda_\gamma$, $\Lambda_e$ parameterise the coupling. They can also be expressed in terms of the dimensionless parameters $d_e$ and $d_{m_e}$, with $d_{e,m_e}=M_\mathrm{Pl}/(\sqrt{4\pi}\Lambda_{\gamma,e})$, where $M_\mathrm{Pl}$ is the Planck mass. The terms in \Cref{eq:L_int} cause effective changes of the fine structure constant $\alpha$ and the effective rest mass $m_e$~\cite{derevianko2014,arvanitaki2015}.
These changes in turn modify the lattice spacing and electronic modes of solids, driving modulations of size $l$ and refractive index $n$:
\begin{align}\label{eq:delta_l}
    \frac{\delta_l}{l}&= - \left( \frac{\delta_{\alpha}}{\alpha} +   \frac{\delta_{m_e}}{m_e}\right),\\
    \frac{\delta_n}{n}&= - 5\cdot 10^{-3}\left(2 \frac{\delta_{\alpha}}{\alpha} +   \frac{\delta_{m_e}}{m_e}\right), \label{eq:delta_n}
\end{align}
where $\delta_x$ denotes a change of the parameter $x$: ${x\rightarrow x + \delta_x}$. Eqs. \ref{eq:delta_l}, \ref{eq:delta_n} hold in the adiabatic limit, which applies for solids with a mechanical resonance frequency much higher than the driving frequency $\omega_\phi$~\cite{geraci2019,arvanitaki2016,grote2019}. 

\begin{figure}[b]
    \centering
    \includegraphics[scale=0.3]{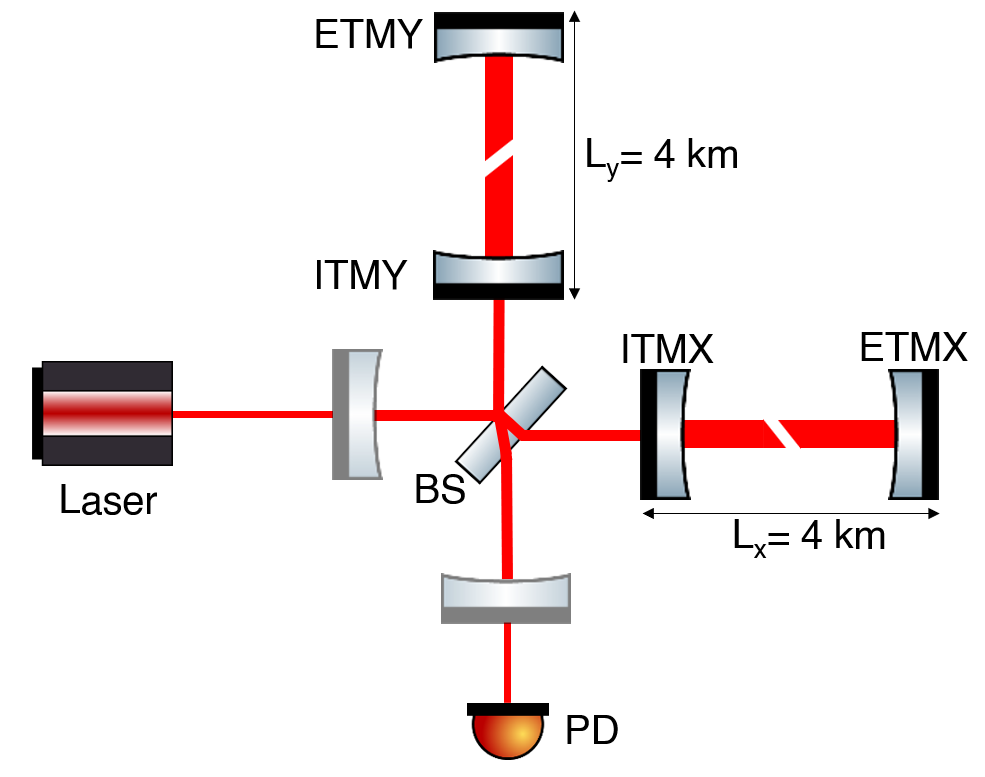}
    \caption{Simplified optical layout of a LIGO-type interferometer. A beamsplitter BS splits a laser beam into two long arms that contain Fabry-Pérot cavities composed of test masses ITMX/Y and ETMX/Y, respectively. The interferometric output is read by a photodetector PD.}
    \label{fig:opt}
\end{figure}

\noindent In the LIGO interferometers, light from a laser source impinges on a beamsplitter (BS) and splits into two orthogonal arms, each containing a Fabry-Pérot cavity (comprised of two mirrors, referred to as test masses) to increase the effective optical path length and optical power of the arms. A sketch of this optical layout can be seen in \Cref{fig:opt}.
While all components of the interferometers can be affected by DM, the BS has been identified as a dominant coupling element for scalar field dark matter, as argued in~\cite{grote2019}. This is because the ``splitting'' effect occurs on one surface of the BS and not at its centre of mass, see \Cref{fig:opt}. This results in DM causing a path length difference between the arms.

\noindent For a BS of thickness $t_B$ and index of refraction $n$, one expects from \Cref{eq:osc_field,eq:delta_l} a length change to be produced in the LIGO interferometers~\cite{grote2019}:
\begin{equation}\label{eq:full_signal}
    \delta (L_x - L_y) \approx \left(\frac{1}{\Lambda_\gamma} + \frac{1}{\Lambda_e} \right)\cdot\frac{n\,t_B\,\hbar\,\sqrt{2\,\rho_{\mathrm{local}}}}{m_\phi\, c},
\end{equation}
where $\delta (L_x - L_y)$ is the optical path difference between both arms and we have neglected the contribution of the refractive index changes to the signal, as it is more than two orders of magnitude smaller than that of the size changes. 
In this work, for the first time, we also take into account the contribution of the four arm test masses, which predominantly produce a signal by changing the optical path lengths within the arms. While this effect mostly cancels out if the test masses have identical thicknesses, as pointed out in \cite{grote2019}, we find that the real small thickness differences between LIGO's test masses lead to non-negligible additions to the BS-induced coupling. 

Length fluctuations, such as those caused by the BS and the test mass couplings, are transduced by the optical interferometric setup into signals on the photodetector (see \Cref{fig:opt}). This conversion can be represented by so-called \textit{transfer functions}, which describe how the interferometer responds to signals of different frequencies. In LIGO, the photodetector signal $I_\text{PD}(\omega)$ is calibrated to GW-induced strain $h(\omega)$ according to:
\begin{equation} \label{eq:readout}
    h(\omega) = \frac{{I_\text{PD}(\omega)}}{{L \, \text{T}_{\rm GW}(\omega)}\,e^{i \phi _{\rm GW}}},
\end{equation}
where $L$ is the arm length of the interferometer ($\approx$\SI{4}{\kilo\meter}), and T$_{\rm{GW}}$ is the optical transfer function from GW-induced strain (with phase $\phi_{\rm GW}$) to photodetector signal. However, to search for the expansion effect of scalar field dark matter, we are interested in a different type of strain that corresponds to thickness changes of the optical components of the interferometer, as described by \Cref{eq:full_signal} for the beamsplitter. To also take into account the effect of the arm test masses, we define:
\begin{equation} \label{eq:mirror_thicknesses}
   t_\text{M} = \left(t_{\text{ETMY}} + t_{\text{ITMY}}\right) - \left(t_{\text{ETMX}} + t_{\text{ITMX}}\right),
\end{equation}
where $t_{\text{ETMY}}$, $t_{\text{ITMY}}$, $t_{\text{ETMX}}$, and $t_{\text{ITMX}}$ represent the mirror thicknesses for the End Test Mass (ETM) and Input Test Mass (ITM) in the Y- and X- interferometer arms, respectively. The thickness variations of the optics under the effect of DM, \textit{i.e.} the DM-induced strain $s_{\rm DM}(\omega)$, can then be expressed as:
\begin{equation}\label{eq:DM_strain_0}
    s_{\rm DM}(\omega) = \frac{I_{\text{PD}}(\omega)}{|n\,t_\text{B} \text{T}_{\rm B} e^{i \phi _{\rm B}} + t_\text{M} \text{T}_{\rm M} e^{i \phi _{\rm M}}|},
\end{equation}
where $\text{T}_{\rm B}$ and $\text{T}_{\rm M}$ are the transfer functions corresponding to the beamsplitter and test mass effects, respectively, and $\phi_{\rm B}$ and $\phi_{\rm M}$ are the phases of those transfer functions. Since we have access to the GW-induced strain $h(\omega)$ only (and not $I_\text{PD}(\omega)$), we express the DM-induced strain as:
\begin{equation}\label{eq:DM_strain}
    s_{\rm DM}(\omega) = h(\omega)\cdot\left|\frac{L\,\text{T}_{\rm GW}\, e^{i \phi _{\rm GW}}}{n\,t_\text{B} \text{T}_{\rm B} e^{i \phi _{\rm B}} + t_\text{M} \text{T}_{\rm M} e^{i \phi _{\rm M}}}\right|.
\end{equation}

\noindent Finally, we can express $h(\omega)$ in relation to the coupling constants $\Lambda_\gamma$ and $\Lambda_e$:

\begin{align} \label{eq:full_signal1}
h(\omega)&\cdot A_{\rm cal}(\omega) \approx \left(\frac{1}{\Lambda_\gamma} + \frac{1}{\Lambda_e}\right) \cdot\left(\frac{\hbar\,\sqrt{2\,\rho_{\mathrm{local}}}}{m_\phi\, c}\right), \\
\text{where} & \nonumber\\
A_{\text{cal}} = &\frac{{\rm T}_{\text{GW}}\cdot L / (n\,t_\text{B} {\rm T}_{\text{B}})}{\sqrt{1 + 2\psi \cos(\phi_{\rm B} - \phi_{\rm M}) + \psi^2}}; \psi=\frac{t_\text{M} {\rm T}_{\text{M}}}{n\,t_\text{B} {\rm T}_{\text{B}}}.  \nonumber
\end{align}

We derive the required GW and DM transfer functions considering both the BS and test mass effects using a simulation-based approach. While the underlying principles of the transfer function calculation are rigorously understood analytically, see~\cite{Mizuno1993}, this approach is better suited for handling the many complexities specific to individual optical setups. We achieve this using the Finesse~\cite{finesse} software package, which was designed to model optical-interferometric systems in the frequency domain and has been widely corroborated experimentally.
We note that the simulation also takes into account any effects stemming from the light travel time, as have been pointed out in \cite{Morisaki2021}.
We also considered the phase difference between end test masses that is caused by the finite de Broglie wavelength of the DM field. This increases the total transfer function's magnitude by about $5\%$ at \SI{5}{\kilo\hertz} when averaged over Earth's sky coverage. Given its negligible impact on our results below, when compared to statistical uncertainty, we disregard this effect here.

\begin{figure}[h]
    \centering
    \includegraphics[scale=0.5]{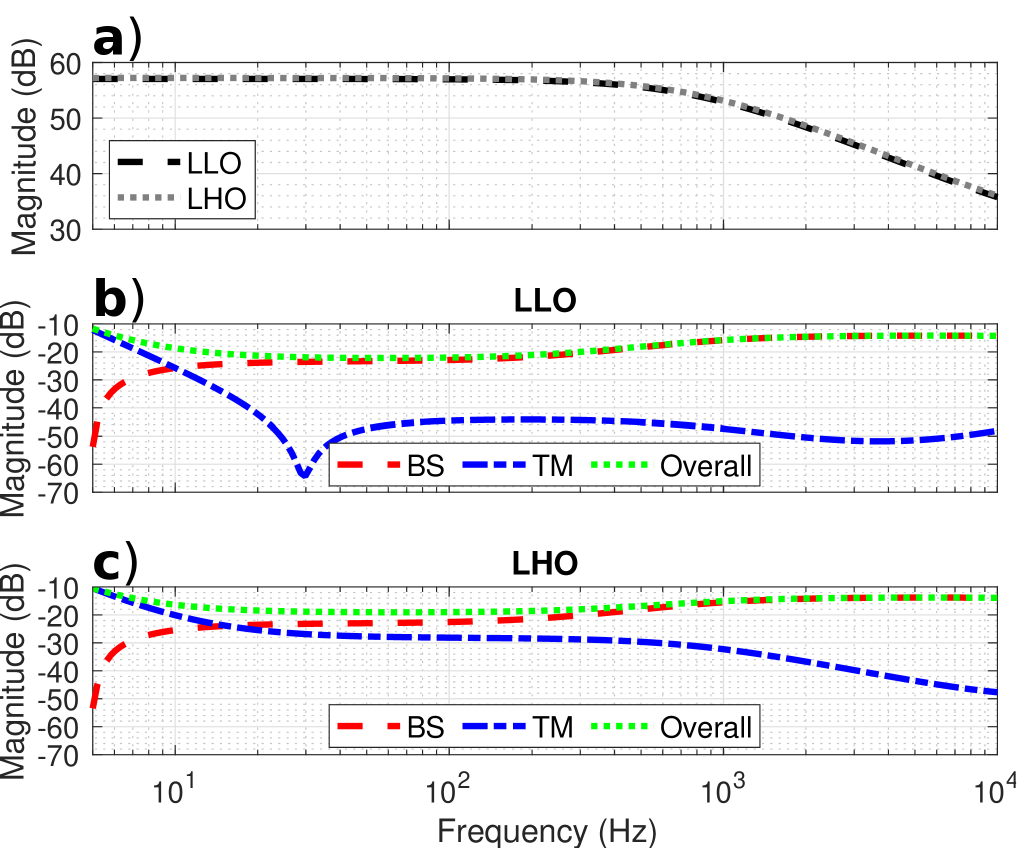}
    \caption{Simulated transfer functions as a function of frequency. \textbf{a}) $T_{\text{GW}}$ for both interefometers, \textbf{b}) and \textbf{c}): $T_{\text{M}}$ for the test-mass effect (dashed line) and $T_{\text{S}}$ for the BS effect (dot-dashed line), see text, and their in-phase combination (dotted line) for LLO and LHO, respectively.}
    \label{fig:TF1}
\end{figure}

\noindent The obtained transfer functions, as well as individual results for BS and test mass effects are shown in \Cref{fig:TF1} for the LIGO Livingston (LLO) and Hanford (LHO) observatories, respectively. The most relevant optical data for this simulation are listed in \Cref{tab:Fin_Para}.
\begin{table}[ht]
    \centering
    \begin{tabular}{l c c c c} 
        \hline
        \hline
                  & \multicolumn{2}{c}{LHO} & \multicolumn{2}{c}{LLO} \\
                  & Thickness & Transm. & Thickness & Transm. \\
                  & (mm)      & (\%)    & (mm)      & (\%)\\
        \hline
        BS	      & 60.41          & 50           & 59.88          & 50 \\
        ITMX      & 199.763   & 1.5          & 199.960   & 1.48 \\
        ETMX      & 199.846   & 3.9e-4         & 199.245   & 7.1e-4 \\
        ITMY      & 199.904   & 1.5          & 199.290   & 1.48 \\
        ETMY      & 199.792   & 3.8e-4         & 199.954   & 7.6e-4 \\
        \hline
        \hline
    \end{tabular}
    \caption[]{Relevant thickness and transmission values~\cite{opt_values} of relevant optical components in LHO and LLO, as used for our transfer function simulations. See \Cref{fig:opt} for an overview of the different components and their meaning.}
    \label{tab:Fin_Para}
\end{table}
\noindent As can be seen from the middle and bottom panels in \Cref{fig:TF1}, the influence of the BS in both detectors becomes dominant at \SIrange{10}{20}{\hertz}. The differences between the detectors is caused by small differences in test mass thicknesses.

\subsection*{Logarithmic spectral analysis}
Given the frequency-dependence of the expected DM signal, with ${\Delta\omega_{\text{obs}}/\omega_{\text{obs}}\sim 10^{-6}}$, maximising signal to noise ratio in our analysis implies a bin spacing in frequency space with the same constant width-to-frequency ratio as that expected from the signal~\cite{dooley2016,vermeulen2021a}. The dataset used in this paper is from LIGO's third observing run~\cite{abbott2023}. We use 40 segments of data that are at least \SI{28}{\hour} in length, to ensure integration over at least one coherence time at our lower frequency bound (\SI{10}{\hertz}), with a total of about \SI{1500}{\hour} of data sampled at \SI{16}{\kilo\hertz}.
The calculation of Discrete Fourier Transforms (DFT) we thus require presents a unique technical challenge, as it explores unprecedented frequencies (below \SI{50}{\hertz}) for this kind of analysis.
It for example needs to use amounts of data ($\mathcal{O}(100$ GB)) exceeding typical memory capacities. More importantly, the cost of this calculation scales as $\mathcal{O}(N^2)$, where $N$ is the number of data points, leading to a prohibitively expensive regime. While methods exist to accelerate logarithmic DFT calculations~\cite{brown1992,holighaus2013}, none reach the Fast Fourier Transforms' (FFT) speed, and no existing package satisfies this analysis' requirements (including memory).
The use of logarithmic bin spacing precludes the use of the FFT due to the resultant frequency-dependent terms and variable data points in the DFT calculations. However, we observed that this frequency-dependence was relatively weak, allowing us to implement small approximations to modify the DFT. This adjustment allowed for the effective use of FFTs, significantly enhancing computational efficiency.
For details about this calculation and the aforementioned software requirements see \cite{gotteletal.2023}. Overall, we achieve a speed-up factor of $\mathcal{O}(10^4)$ with negligible impact on the results.

Since the effect of DM on the detector cannot be ``turned off'', it is necessary to build a background model that is resistant to the influence of existing peaks in the data (see~\cite{derevianko2018}). This is done, after having calculated the PSD, by implementing a recursive procedure making use of spline fits similar to that used by LIGO calibration~\cite{sun2020}. In each iteration, bins containing identified peaks are removed and the fits are repeated on the ``cleaned'' data, until the solutions converge.
The method was validated by varying the degrees of freedom of the aforementioned approximation.

\noindent We find empirically that the residuals of the background model with respect to the log of the observed PSD are well described by a skew-normal distribution. While the parameters of said distribution vary over frequency, this variation is small: the following analysis is thus performed in chunks of 10,000 frequency bins in which the distribution parameters can safely be viewed as constant. We use a likelihood-based analysis in order to combine data from different segments (in time) and from the different interferometers. The likelihood is defined as:
\begin{equation}
    \mathcal{L}(\mu, \vec{\theta}, \tilde{\vec{\theta}}) = \sum_{\text{seg,ifo}}\sum_{j=0} \log f_{seg}\left(g_{j,\text{seg,ifo}}(\mu, \vec{\theta}), \tilde{\vec{\theta}}\right),
\end{equation}
where the sums are over data segments and frequency, respectively, $\mu$ is proportional to the amplitude of the DM peak, $f_\text{seg}$ is a skew normal distribution with parameters held by $\tilde{\vec{\theta}}$, the subscript $j$ denotes the frequency bin index, seg the data segment, and ifo the corresponding interferometer. $g_{\text{seg,ifo}}$ represents the residuals between the data and the expected background, with a term to allow for DM effects:
\begin{equation}
     g_{j,\text{seg,ifo}} = \log Y_\text{seg} - \log\left(e^{bkg_\text{seg}} + \mu\cdot\beta_\text{ifo}\right),
\end{equation}
where $Y(\omega)$ is the PSD data based on $h(\omega)$, $bkg(\omega)$ refers to the fitted background shape in log space, and $\beta_\text{ifo}(\omega)$ is an interferometer-specific calibration term based on \Cref{eq:full_signal1}. Frequency-dependence throughout the equation is left implicit for simplicity. The parameter $\mu = \Lambda_i^{-2}$ was chosen because, as can be seen in \Cref{eq:full_signal1}, it is not possible to differentiate between a non-zero $\Lambda_{\gamma}^{-1}$ or $\Lambda_{e}^{-1}$. $\Lambda_i^{-1}$ can thus be interpreted as being either one of the two coupling constants, under the assumption that the other is null.
Although the proximity of the interferometers could allow us to exploit coherence effects, the abundance of non-coincident data segments in our dataset prompted us to disregard this method, as we estimated that it would lead to only a roughly 10\% improvement in our results.

\noindent In this framework, finding a DM signal is thus equivalent to rejecting the hypothesis that $\mu = 0$. Additionally, in order to correctly make use of the fact that physically, $\mu$ cannot be negative, we use the profile-likelihood-ratio based test statistic $q_0$, as described in eq. (12) in~\cite{cowan2011}, and corresponding asymptotic methods, to search for a positive signal.

\noindent Conversely, in order to calculate the upper limit on $\Lambda_i^{-1}$, the complementary test-statistic $\Tilde{q_{\mu}}$ as described in eq. (16) in~\cite{cowan2011}, was used.
This test-statistic correctly accounts for cases where the value of $\mu$ maximizing the likelihood is greater than the hypothesized value, ensuring that upward noise fluctuations are not considered as less compatible with a given upper limit.

\begin{figure}[hb]
    \centering
    \includegraphics[width=.5\textwidth]{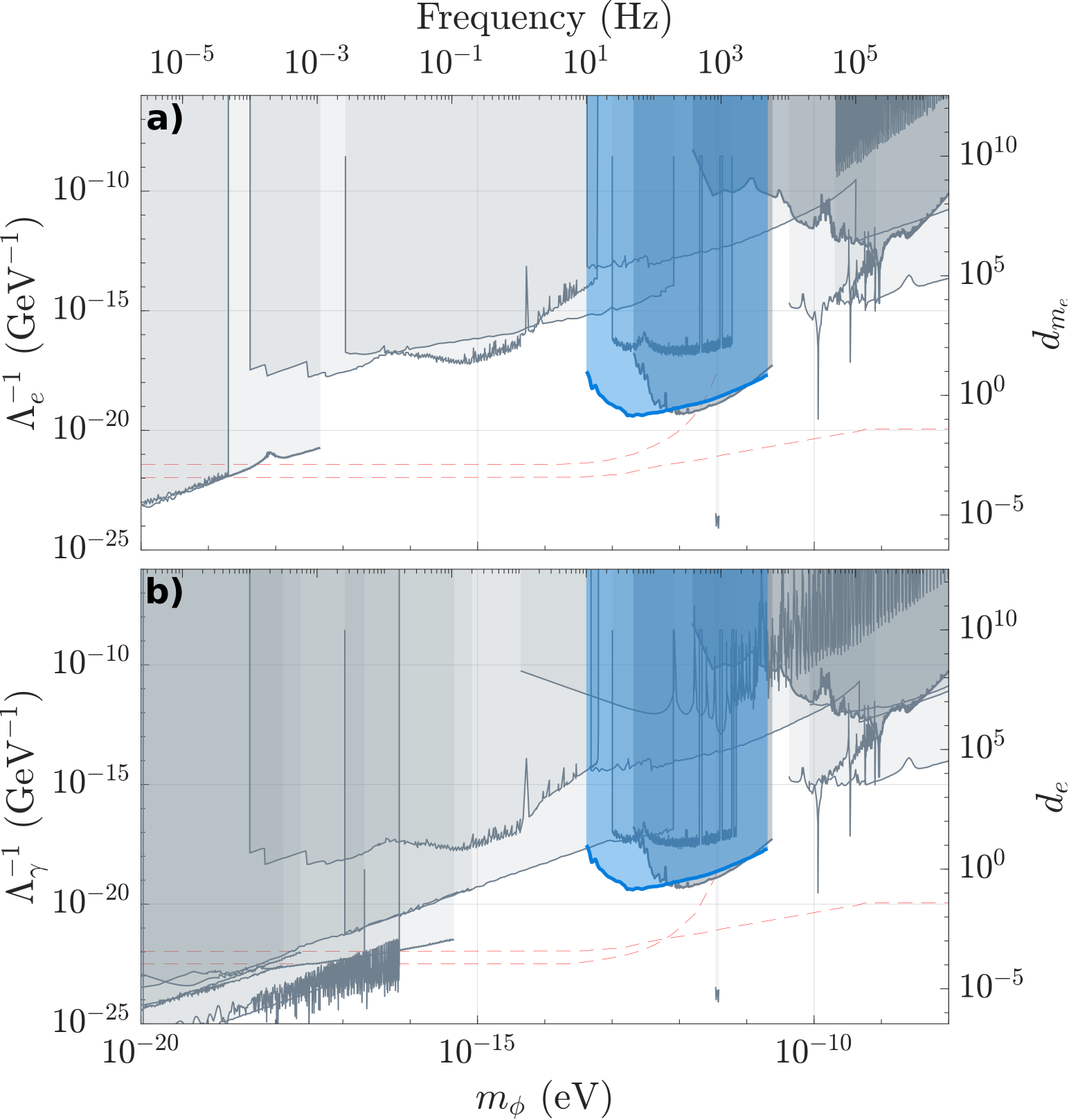}
    \caption{Upper limit on $\Lambda_i^{-1}$ (95\% C.L.) as a function of frequency. \textbf{a}) and \textbf{b}) depict our results in the context of other experimental results on $\Lambda_e$, $\Lambda_{\gamma}$, respectively. Our results are shown by the thick blue line, constraints from direct experimental searches for DM~\cite{vermeulen2021a,Aiello2022,aharony_constraining_2021,savalle_searching_2021,kennedy_precision_2020,antypas_scalar_2019,antypas_probing_2021,tretiak_improved_2022,oswald_search_2022,campbell_searching_2021,beloy_frequency_2021,zhang_search_2023,Fukusumi2023,sherrill2023} are shown in thin grey, and constraints from searches for `fifth 
    forces'~\cite{berge_microscopes_2021,hees_violation_2018} are depicted by the dashed red lines. Our results were smoothed for visual purposes.}
    \label{fig:results}
\end{figure}

This approach enabled the search for DM signals by identifying local excesses in the $q_0$ value across different frequency bins. A $5\sigma$ threshold, corrected for the \textit{look-elsewhere-effect}, resulted in 349 candidates, which was reduced to 159 by associating neighbouring over-threshold bins to single candidates. Finally, since the reconstructed amplitude of DM should not vary much over time, the consistency of the results was further probed with a t=5 threshold on a student-t test comparing results from different segment combinations. A final cut on the remaining 42 candidates was then set on requiring that both interferometers have results that are significantly different from zero. Faced with the lack of surviving candidates, our upper limits can be seen in the context of other measurements in \Cref{fig:results} for $\Lambda_e^{-1}$ and $\Lambda_{\gamma}^{-1}$, respectively.

\noindent These results assume a local dark matter density $\rho_\mathrm{CDM}=0.4$~GeV/cm$^3$ (as in \cite{freese2013} for the standard smooth DM halo model). Models in which DM forms a relaxion halo \cite{graham2015,kolb1993} predict local DM overdensities of up to $\rho_\mathrm{RH}/\rho_\mathrm{CDM}\leq 10^{16}$ \cite{savalle2019}. Our results impose significantly more stringent constraints on the coupling constants for higher assumed values of the DM density $\rho_\mathrm{A}>\rho_\mathrm{CDM}$: the constraint becomes more stringent by a factor $(\rho_\mathrm{A}/\rho_\mathrm{CDM})^{1/2}$ (see Eq.~\ref{eq:full_signal}).

\noindent Our limits represent a several order of magnitude improvement on other direct searches in a band from \SIrange{10}{180}{\hertz} (roughly \SIrange{5e-14}{1e-12}{\electronvolt}). The main limiting factor being detector noise, we expect those results to be improved greatly in future LIGO runs and with future gravitational wave detectors. We emphasise in particular that the results could also be improved drastically by increasing mirror thickness differences in the interferometer arms, for which this study paves the way.

\section*{Acknowledgements} \label{sec:acknowledgements}
The authors are grateful for support from
the Science and Technology Facilities Council (STFC), grants
ST/T006331/1 and ST/W006456/1 for the Quantum Technologies for Fundamental Physics program, as well as ST/I006285/1, and ST/L000946/1, and the Leverhulme Trust, grant RPG-2019-022.
This work was supported in part by Oracle Cloud credits and related resources provided by the Oracle Corporation.
This research has made use of data or software obtained from the Gravitational Wave Open Science Center (gwosc.org), a service of the LIGO Scientific Collaboration, the Virgo Collaboration, and KAGRA. This material is based upon work supported by NSF's LIGO Laboratory which is a major facility fully funded by the National Science Foundation, as well as the Science and Technology Facilities Council (STFC) of the United Kingdom, the Max-Planck-Society (MPS), and the State of Niedersachsen/Germany for support of the construction of Advanced LIGO and construction and operation of the GEO600 detector. Additional support for Advanced LIGO was provided by the Australian Research Council. Virgo is funded, through the European Gravitational Observatory (EGO), by the French Centre National de Recherche Scientifique (CNRS), the Italian Istituto Nazionale di Fisica Nucleare (INFN) and the Dutch Nikhef, with contributions by institutions from Belgium, Germany, Greece, Hungary, Ireland, Japan, Monaco, Poland, Portugal, Spain. KAGRA is supported by Ministry of Education, Culture, Sports, Science and Technology (MEXT), Japan Society for the Promotion of Science (JSPS) in Japan; National Research Foundation (NRF) and Ministry of Science and ICT (MSIT) in Korea; Academia Sinica (AS) and National Science and Technology Council (NSTC) in Taiwan.
This document has been assigned LIGO document number LIGO-P2400010.
\vfill


\end{document}